\documentstyle[aps,pre]{revtex}
\begin{document}
\draft
\title{Permeability of self-affine rough fractures}
\author{German Drazer and Joel Koplik}
\address{Benjamin Levich Institute and Department of Physics,
City College of the City University of New York, New York, NY 10031}
\date{\today}
\maketitle
\begin{abstract}
The permeability of two-dimensional fractures with self-affine fractal
roughness is studied via analytic arguments and numerical simulations.
The limit where the roughness amplitude is small compared with average 
fracture aperture is analyzed by a perturbation method, while in the 
opposite
case of narrow aperture, we use heuristic arguments based on lubrication
theory.  Numerical simulations, using the lattice Boltzmann method, are
used to examine the complete range of aperture sizes, and confirm the 
analytic arguments.   
\end{abstract}
\section{Introduction}
\label{introduction}

The transport of fluids through geological media such as hydrocarbon and
water reservoirs involves a combination of flow through the
microscopic pore space of the rock itself and flows through macroscopic
channels such as fractures.  The first case is relatively well understood,
at least in principle \cite{dullien,sahimi93}, 
by means of models which treat the 
pore space as a random network, and then use effective medium or 
percolation
concepts for the transport.  Fractures are typically modeled as simple
straight-sided channels, with a cubic relation between fluid flux and
average aperture, and a challenging problem is to understand the dynamics
of flow in a macroscopic fracture network \cite{adler99,bear93,sahimi95,nas}.  
Typically
the surface of a single fracture appears fairly smooth, aside from some
small-scale superficially random roughness, and Poiseuille flow in a straight
channel is the obvious model for fluid flow.  However, more careful analysis
\cite{bouchaud97} shows that common geological fractures in fact have
correlated, self-affine fractal surfaces.  The roughness exponent, whose
precise definition is recalled below, is usually 
found \cite{bouchaud90,plouraboue96} to be close to 0.8, surprisingly 
insensitive to the material and the fracturing process.  (Other values 
may arise from intergranular effects \cite{boffa}.)

The aim of this paper is to study the permeability of single self-affine 
two-dimensional fractures.  We shall see that it is possible to obtain general 
analytic expressions in two limiting cases, where the roughness associated 
with the fracture surface is either small or large compared to the mean 
aperture.  The theoretical predictions will be supported by numerical 
simulations, using the lattice Boltzmann method, which of course can 
address the case of intermediate sized roughness as well.   In subsequent 
work, we will consider the fully three dimensional case, and go on to 
consider more general transport processes in fractures, involving both 
passive tracers and 
finite-sized suspended particles.  Complimentary experimental studies on 
laboratory samples of fractured rock are in progress elsewhere \cite{jph}.

To the extent that the fluctuations in the height are small compared to the 
aperture width, the effect on permeability is modest, and we will obtain 
small corrections to the usual cubic law, but the effects on other transport 
processes are much more significant.  For example, passive tracer 
dispersion is very sensitive to the presence of any slow zones in the velocity 
field, and even uncorrelated roughness leads to significant long-time tails 
\cite{ippolito}.   Similarly, the deposition and erosion of solid particles from a 
surface is sensitive to the local shear stress, which in turn can change 
dramatically as a surface roughens.

In the opposite extreme of a very narrow fracture,
the flow field changes qualitatively as the fluid winds through a 
highly irregular channel.  Aside from the challenge of estimating the 
permeability of difficult geometry in terms of its statistical characteristics,
there are further effects arising from the fact that the two sides of the 
fracture originate from a single crack.  In the fractal case, the spatial 
correlations between the two sides of a  fracture lead to velocity field 
correlations, which again strongly affect tracer motion 
\cite{roux93,roux98,plouraboue98}.  The latter references made use of
 a very approximate velocity field, obtained in the lubrication limit, and
one of the motivations of this paper is to examine the validity of the
lubrication approximation for the velocity field.  

The organization of this paper is as follows.  We first recall some basic 
facts about self-affine surfaces, and give the algorithm used to generate them 
numerically,  and also explain the lattice Boltzmann method used for the 
numerical simulations.  We then consider the case of fractures with a wide
average aperture, obtain a perturbative estimate for the permeability, and 
test it numerically. We then turn to narrow fractures, first in the case where 
the two sides are simply displaced normally to the mean fracture plane, and 
secondly when there is a lateral shift as well. Finally, we summarize the results, 
and indicate the next steps 


\section{Preliminaries}


\subsection{Self-affine roughness}
\label{self-affine}

We briefly review the mathematical characterization of self-affinity,
and its implementation in this paper.
We consider a rock surface without overhangs, whose height is given by a
by a single-valued function $z(x,y)$, where the coordinates $x$ and $y$ lie
in the mean plane of the fracture.  Self-affine surfaces \cite{feder} 
display scale invariance with different dilation ratios along different 
spatial directions, remaining unchanged under the rescaling
\begin{equation}
x \rightarrow \lambda_1 x \qquad
y \rightarrow \lambda_2 y \qquad
z \rightarrow \lambda_3 z
\end{equation}
Here we consider disordered media, so these scaling laws apply only in an 
ensemble or spatial average sense.  Experiment indicates that for many
materials isotropy can be assumed in the mean plane, implying that there
is only one non-trivial exponent relating the dilation ratio
in the mean plane to the scaling in the perpendicular direction, {\em i.e.},
$\lambda_1=\lambda_2\equiv\lambda$, and 
$\lambda_3=\lambda^\zeta$, with
\begin{equation}
\label{scaling}
z(x,y) = \lambda^{-\zeta} z(\lambda x, \lambda y)
\end{equation}
where $\zeta$ is the roughness or Hurst exponent \cite{mandelbrot}. 

We assume that the process of fracturing the rock is ``clean'', in the sense
that the rock breaks so as separate along one single valued surface without
subsequent deformation and without producing 
loose interstitial material.  We shall emphasize two complimentary 
limiting situations, in which the two surfaces are either well separated 
with respect to each other, or very close to each other.  If $L$ is the 
lateral size of the fracture in the mean plane, the typical range of the
fluctuations (the difference between the maximum and minimum values of 
$z$) scales as $R\sim L^\zeta$. If $h$ is the mean width of the fracture, 
with $h \ll L$, then
the two cases correspond to $R \ll h$ and $R \gg h$, respectively.

The local aperture of the crack is the key 
parameter determining the fracture permeability, and three different cases 
will be addressed.  For large aperture, the roughness associated with the
two sides of the fracture act independently, and it suffices to consider a
channel with one rough and one smooth boundary.  In the case of a narrow
fracture, the correlation between the two sides is a key feature, and we
consider two possibilities.  We first study fractures where the upper 
surface has been simply translated a distance $h$ normal to the mean plane,
so that the local aperture $a(x,y)$ equals the constant $h$. 
Alternatively, the two surfaces may have a relative lateral displacement $d$
in the mean fracture plane, accompanied by a displacement $h$ in the 
perpendicular direction, so that the two surfaces do not overlap.
In this case the local aperture is given by the random variable.
\begin{equation}
\label{aperture}
a_d(x,y) = z(x+d,y) - z(x,y) + h
\end{equation}
It turns out \cite{plouraboue95} that $d$ is the lateral correlation length 
for fluctuations in the aperture, in the sense that $a_d(x,y)$ and 
$a_d(x+\Delta x,y)$ decorrelate for $\Delta x\gg d$.

In this paper we restrict ourselves to the two dimensional case 
where the surface is invariant in the $y$ direction, $z(x,y)=z(x)$, 
and the flow is forced in the $x$ direction by a constant pressure drop. 
In a subsequent paper we will extend these calculations to fully three
dimensional fractures, but it is convenient, both conceptually and in the
numerical simulations, to regard the system as having a translationally
invariant third dimension.


\subsection{Numerical Methods}
\label{numerical}

Our aim is to study various aspects of the fracture permeability
which are sensitive to the fracture roughness.  The first ingredient 
is the height function $z(x,y)$, a statistically self-affine surface
with periodic boundary conditions.  The periodicity is not a physically
essential ingredient here, but has some calculational advantages in 
alleviating finite-size effects.  The surface is generated by a Fourier 
synthesis method, based on power-law filtering of arrays of 
independent random numbers \cite{voss85}. The random numbers
are generated using a Gaussian distribution, and then 
modulated by an appropriate power law.  If $\tilde{Z}(k)$ is the Fourier
transform  of the initial Gaussian random array, then the Fourier
transform of the surface elevation is chosen to be
\begin{equation}
\label{spectrum}
Z(k) = k^{-\zeta-1/2} \tilde{Z}(k)
\end{equation}
(The 1/2 is appropriate for the self-affine {\em curve} used here, and 
would be 1 for a real rock {\em surface}.)

In most cases the 
roughness exponent is chosen as the often-observed value $\zeta=0.8$.
The amplitude of the roughness can be expressed in terms of  
variance of the height distribution over the full range, 
\begin{equation} 
\sigma^2 = \langle ( z(x) - \langle z(x) \rangle )^2 \rangle
= \frac{1}{L} \int_0^L ( z(x) - \langle z(x) \rangle )^2 dx,
\end{equation}
which in this case is $\sigma^2=0.49\pm 0.06$ for $L=256$.  Alternatively, 
the amplitude is given indirectly in terms of the range of the fluctuations 
as $R = R_1 L^{\zeta}$, with $R_1=1.66\pm 0.4$.   In these equations, 
and in the rest of the paper, the unit of length is that of the spacing 
between lattice points in the numerical calculations.

Since we consider fluid flow in a highly irregular geometry, and eventually
hope to consider dispersion and the evolution of the solid surface due to 
particle
transport, the lattice Boltzmann (LB) method \cite{rothman} is particularly 
convenient.   In this algorithm, fictitious particles move between
neighboring sites on a lattice, with suitable collision rules, and the 
boundary of the flow domain is simply a surface of sites where the rule 
is modified in some way to keep the particles out.  We use the 
Face Centered Hypercubic (FCHC-projected) version of the LB model,
with a cubic lattice in 3 dimensions and 19 velocities
(D3Q19 in the terminology of \cite{qian}). The collision operator is 
approximated by a single relaxation time, the Bhatnagar-Gross-Krook 
model\cite{BGK}, and the local equilibrium distribution given by 
\cite{qian} is used. This pseudo-equilibrium distribution locally
preserves mass and momentum values, and is formulated specifically to 
recover the Navier-Stokes equation at large length and time scales.
For the no-slip solid boundaries, we use the simplest implementation of 
particle exclusion -- the bounce-back rule, in which a particle incident on
the boundary reverses its direction.  
Periodic boundary conditions are used for 
the inflow and outflow surfaces.  A constant pressure gradient forcing the 
fluid is added in the $x$ direction, while the gap between surfaces extends 
over $z$, and the geometry is taken to be translation invariant in $y$. 


\section{Small surface deviations}

In this section we consider the case of a channel with one fractally-rough
wall, in the limit where the mean width is large compared to the amplitude
of the roughness.  We (re-)derive an elegant general result
for the permeability, and then an exact variant which gives the permeability
of a perturbed pore space.  If the pore space perturbation is weak, the
leading order correction is easy to evaluate and, when compared to 
numerical simulation, is seen to provide a reasonably accurate estimate of 
the permeability.

\subsection{Integral representations}
\label{integral}   

We begin by deriving an integral representation for the permeability, 
and then considering a perturbation in the boundary shape.  Although we 
suspect that these results are known to many, we are not aware of an earlier 
publication containing them, although the electrical analog of the 
integral representation is in the literature \cite{bergman89}. 

Consider a rectangular volume of a porous medium, periodic in all
directions, and the following surface integral 
\begin{equation}
\label{surface}
I=\int_S d\vec{s} \cdot \Big[\vec{u}(\vec{r})~p(\vec{r})\Big] 
\end{equation}
Here, $\vec{u}$ and $p$ are the velocity and pressure fields of a fluid 
which completely fills the pore space, assumed to satisfy the Stokes
equations, and the surface $S$ is the complete boundary of the pore space, 
consisting of the inner grain surfaces and the porous regions of the outer
boundary.  The pressure and velocity fields are periodic, except that the
pressure jumps by a constant amount $P$ between two opposite faces in 
one direction, that of the mean flow.  Now the velocity vanishes on the grain 
boundaries by the no-slip
condition, and there is complete cancelation between the opposite faces in
the two completely periodic directions, but in the flow direction the 
remaining two faces of the box combine to give
\begin{equation}
I = P \int_{E} d\vec{s} \cdot \vec{u}(\vec{r}) = -PQ  = -{k P^2 S\over \mu L}
\end{equation}
Here $Q$ is the fluid flux through this end face $E$, and the minus sign 
arises because the outward normal is used in $I$.  The last equality
follows from Darcy's law, where $S$ is the area of end $E$, $k$ is 
the permeability, $\mu$ the fluid viscosity, and $L$ the length of the box. 

On the other hand, applying the divergence theorem to Eq.~\ref{surface},
we obtain:
\begin{equation}
\label{surface1}
I = \int_V dV~ \nabla\cdot(\vec{u}~ p) = \int_V dV~(\vec{u}\cdot\nabla p) 
= \mu \int_V dV~(\vec{u}\cdot\nabla^2\vec{u})=-\mu \int_V 
dV~(\nabla\vec{u})^2
\end{equation}
where the volume $V$ is just the pore space.  We have used first the 
incompressibility of the fluid, second the Stokes equation, and third 
integration by parts. Comparing the two expressions for $I$, we have
\begin{equation}
\label{permeability}
k = \frac{L}{S}~\left(\frac{\mu}{P}\right)^2~\int_V dV~\sum_{i,j} ~
(\partial_i u_j)^2
\end{equation}
The analogous formula for the electrical conductivity $\sigma$,
which follows from Ohm's law and the Laplace equation \cite{bergman89},  is
\begin{equation}
\sigma = \sigma_0~\frac{L}{S}~{1\over\Phi^2}\int_V 
dV~\sum_{i}~(\partial_i \phi)^2
\end{equation}
Here $\sigma_0$ is the conductivity of the pore fluid, $\phi(\vec{r})$ is 
the potential and $\Phi$ its difference across the sample.  Note that both of 
these formulae may also be derived by comparing the energy dissipation rate
computed microscopically in the pore space to the same quantity computed 
using average fields.

We next derive an exact variant of 
the integral expression for the permeability (\ref{permeability})
due to Wilkinson \cite{wilkinson},  which
allows us to implement a perturbation analysis.  Suppose we begin with a 
pore volume $V_0$ and known Stokes equation solutions $p_0$ and 
$\vec{u}_0$, 
and then {\em contract} the volume to $V$, where the exact (but unknown) 
solution is $p=p_0+p_\epsilon$ and 
$\vec{u}=\vec{u}_0+\vec{u}_\epsilon$.
The result to follow is true even if the volume change and alterations in 
the fields are {\em not} small, but are probably only useful in that limit, 
hence the suggestive subscript $\epsilon$.
Substituting into the dissipation integral on the right hand of 
Eq.~\ref{permeability}, and suppressing the summation 
sign for the moment, we obtain
\begin{equation}
\int_V dV ~ (\partial_i u_j)^2 = 
\int_V dV ~\Big[ (\partial_i u_{0,j})^2 + (\partial_i u_{\epsilon,j})^2\Big] 
+ 2~\int_V dV~\partial_i u_{0,j}~\partial_i u_{\epsilon,j}
\nonumber
\end{equation} 
The last integral on the right hand can be rewritten as 
\begin{equation}
\int_V dV ~\Big[\partial_i ( u_{0,j}~\partial_i u_{\epsilon,j}) - 
u_{0,j}~\partial^2_i u_{\epsilon,j} \Big] 
= \int_S dS_i ~ u_{0,j}~\partial_iu_{\epsilon,j} - 
\int_V dV ~ u_{0,j}~\partial^2_iu_{\epsilon,j}.
\nonumber 
\end{equation}
Noting that $u_{0,j} = -u_{\epsilon,j}$ at the surface $S$ due to the 
no-slip condition, and using $\mu\partial^2_iu_{\epsilon,j} = \partial_j
p_\epsilon$, we may rewrite the previous expression as
\begin{equation}
\int_V dV ~\partial_i u_{0,j}~\partial_i u_{\epsilon,j} = 
- \int_V dV \Big[ (\partial_i u_{\epsilon,j})^2 + \mu^{-1}
\vec{u}\cdot\nabla p_{\epsilon} \Big]
\nonumber
\end{equation}
Using incompressibility, the last term in the integrand can be rewritten
$\nabla\cdot(\vec{u}p_\epsilon)$ and converted to a surface integral,
which vanishes if the pressure is held constant on the ends of the porous
medium.  Finally, we obtain
\begin{equation}
\label{perm_final}
k = \frac{L}{S}~\left(\frac{\mu}{P}\right)^2~\sum_{i,j} 
\int_V dV \left[ (\partial_iu_{0,j})^2 - (\partial_iu_{\epsilon,j})^2 \right]
\end{equation}  

This result is exact but not useful.  However, note that if the relative
change in volume is small, $O(\epsilon)$, the second term is second order 
in the small parameter, so that to first order the decrease in permeability 
involves only the integral of the unperturbed velocity over the deleted
region, which is easy to calculate.


\subsection{Dependence of permeability on system size}
\label{size}

In this section we will determine the decrease in permeability due to the 
presence of a self-affine perturbation in the lower surface of a straight 
channel, and in particular its dependence on the size of the channel. 
To use the previous result, we begin with a straight channel of width $h_0$ 
which completely contains the rough-walled one (see Fig.~\ref{fracture}),
where we have Poiseuille flow 
\begin{equation}
\vec{u}_0 = \frac{1}{2\mu}~\frac{P}{L}~z(h_0-z)~\hat{x}
\end{equation}
and an unperturbed permeability $k_0 = h_0^2/12$.  The lower boundary is
shifted by $z(x)>0$, where we assume that that range $R$ of $z$ satisfies
$\epsilon \equiv R/h_0 \ll 1$.
The only surviving velocity derivative is
\begin{equation}
\partial_z u_{0,x}=\frac{1}{2\mu}\frac{P}{L}(h_0-2z)
\nonumber
\end{equation}
so that the dissipation integral is simply
\begin{eqnarray}
\int_{V} dV \sum_{i,j} (\partial_i u_{0,j})^2 &=& 
\int_{0}^{L} dx \int_{0}^{Y} dy  \int_{z(x)}^{h_0} dz 
\left(\partial_z u_{0,x}\right)^2  \\ \nonumber
&=& 
\left(\frac{P}{\mu L}\right)^2 Y
\int_0^Ldx \left\{ \int_{0}^{h_0} dz - \int_{0}^{z(x)} dz  \right\} 
h_0^2 \left(\frac{z}{h_0}-\frac{1}{2}\right)^2 
\nonumber
\end{eqnarray} 
Up to first order in $\epsilon$ this becomes
\begin{eqnarray}
\int_V dV (\partial_iu_{0,j})^2 &\approx& 
\left(\frac{P}{\mu L}\right)^2 Y \left\{
\frac{h_0^3}{12} L - \frac{h_0^2}{4} \int_0^L {z(x)} dx 
\right\} \\ \nonumber
&\approx& 
\left(\frac{P}{\mu }\right)^2 \frac{S}{L}~ k_0
\left\{ 1 - \frac{3}{h_0} \langle z \rangle \right\}
\nonumber
\end{eqnarray}
where $\langle z \rangle$ is the average of $z$ over the channel length $L$,
and the cross-sectional area is $S=h_0Y$.
The mean perturbation $\langle z \rangle$ is computable for a
specific profile, but here we wish to use a statistical characterization
and relate it to the average properties of an ensemble of self-affine
surfaces.  

The mean height of the surface is half the range, and therefore has the 
scaling  $\langle z(x) \rangle = {1\over 2} R_1 L^{\zeta}$.  Substituting,
and replacing the previous equation in Eq.~\ref{permeability} we obtain 
the first-order perturbative correction to the permeability
\begin{equation}
k \approx
k_0 \left[ 1 - {3 \over 2} C_1 \epsilon \right] 
\end{equation}
We have added an adjustable parameter $C_1$, which 
is expected to lie between $1<C_1<2$, to take account of the distinction 
between open volume and flowing volume.  A value $C_1=1$ means
that the fracture is equivalent to a straight channel
of height equal to the mean height of the surface. On the other hand,
a value $C_1=2$ means that the whole region of fluid 
below the maximum excursion of the surface is not contributing
to the permeability and the system is equivalent to
the maximum channel that does not intersect the surface.

Note the lowest order correction to the permeability obtained here 
results from the reduced volume of a fracture compared to that of a 
straight channel enclosing it, and happens to coincide with the lubrication 
approximation.  The approach taken here allows us to gain some insight in 
the drawbacks of this approximation and further improve them.
Using the previous equation we may write the first
order correction to the flow rate,
\begin{equation}
\label{rate}
Q(L) \approx Q_0 \left[ 1 - 
\frac{3 C_1 R_1}{2 h_0} L^{\zeta} ~  \right] 
\end{equation}

To verify this prediction, we consider a pore space
consisting of a cubic lattice with periodic boundaries in $x$ and $y$, one
impermeable wall at $z=h_0$, and a self-affine rough boundary lying above
$z=0$.  We generate an ensemble of 10 self-affine 
surfaces with exponent $\zeta=0.8$, as discussed above.   
In Fig.~\ref{q-vs-size} we display the correction to the flow rate as a 
function of the size of the system $L$.  The straight line 
is a fit to the expected exponent, which does reasonably well except at the 
smallest and largest values of $x$.  The origins of the discrepancies are 
first that the discretization used in the numerical calculation suppresses 
any roughness less than one unit, which is a significant fraction of the 
system for small $L$.   Moreover, the system should be periodic in $L$,
but the numerical algorithm used to generate it assumes a system size that 
is a power of 2.  As usual, there are computational limitations on the sizes 
we can simulate, and for lengths between 32 and the accessible maximum
of 64, we just truncate the periodic surface.  Finally,  
at the largest system sizes, we are really outside the range of validity
of the estimates.  In fact, the second term in the flux in Eq.~\ref{rate} is 
greater than 1/3 for the parameters here when $L\approx 16$.

The fitted value $C_1$ given that $\zeta=0.8$ is
$C_1=1.30 \stackrel{+2.1}{\scriptstyle{-1.2}}$, meaning that a portion
of the fluid does not contributes to convective transport.  Expressed in 
terms of the surface height variance $\sigma^2$ instead of the span range,
the excluded fluid represents $50\%$ of the roughness region.  
The physical consideration absent in the leading order perturbation 
calculation 
is the fact that fluid velocity sharply decays inside depressions 
and corners were resistive eddies are likely to occur \cite{moffat}.  This 
deficiency is associated with the omitted second term in Eq.~\ref{perm_final}, 
{\em i.e.}, the correction to the unperturbed velocity. 
In Fig.~\ref{fracture} we magnify two regions 
close to the lower and upper surfaces, respectively, showing the 
markedly different way the velocity decays towards the surface.

We have also simulated  a system with an alternative 
roughness exponent $\zeta=0.95$, corresponding to a
surface  with more persistent correlations and in a sense a lower fractal 
dimension (the dimension of the intersection of  the surface with a
plane normal to it is $D=2-\zeta$ \cite{feder}).   In this case the
prefactor in the range scaling law is $R_1=0.6\pm 0.12$. 
In Fig.~\ref{q-vs-size} we 
show the numerical results;  again, good agreement with the predicted 
behavior is obtained.  The fit gives 
$C_1=2.04\stackrel{+3.1}{\scriptstyle{-1.8}}$,
suggesting that the low-velocity zones are even more important in this case, 
presumably because the greater degree of correlation enhances them.


\subsection{Permeability decrease due to zero mean surface fluctuations}
\label{zero-mean}

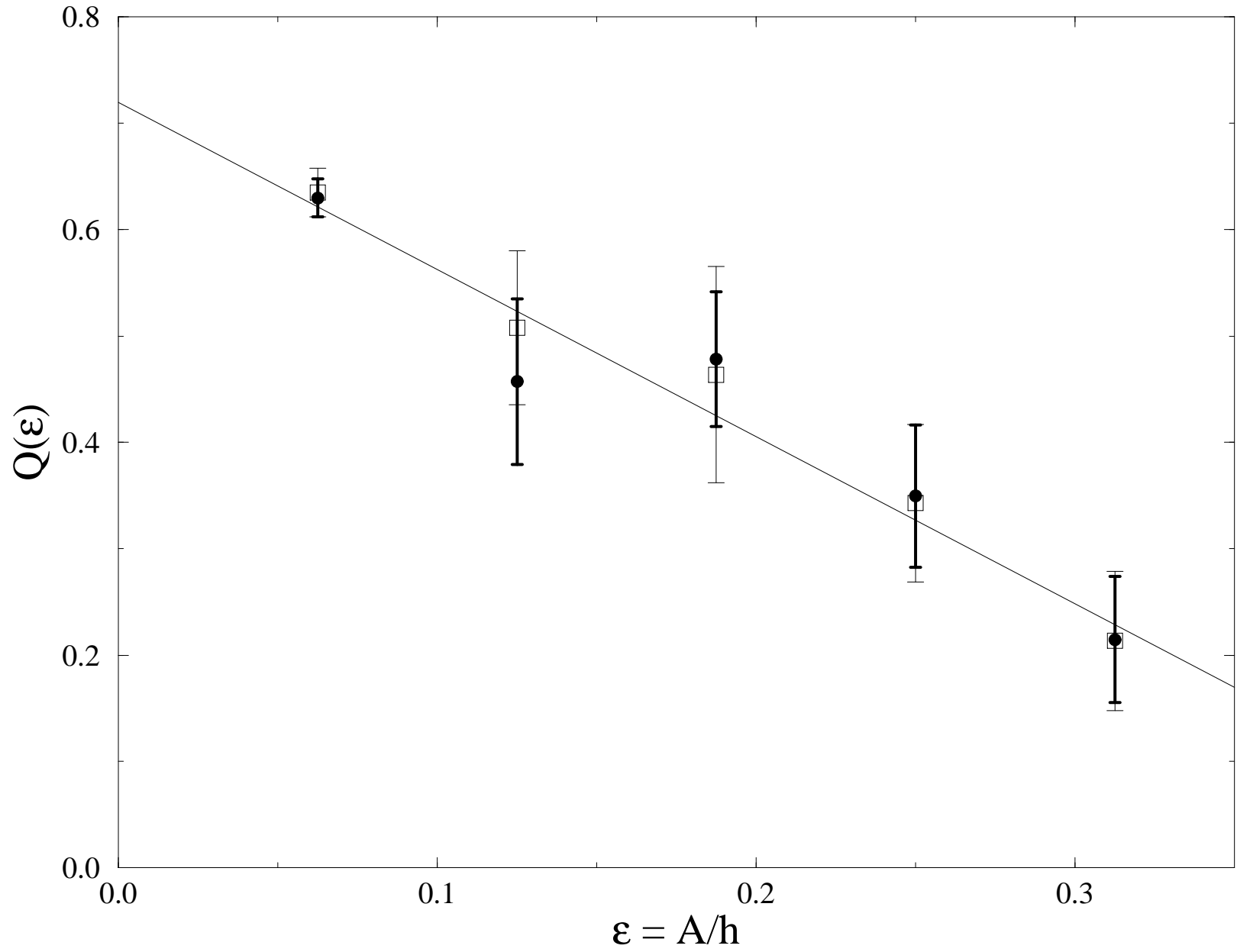
\begin{figure}
\caption{Flow rate as a function of the relative amplitude of the surface
roughness.  Circles and squares correspond to the fully rough and the
small-wavelength filtered surfaces, respectively.}
\label{fig-zero-mean}
\end{figure}

We wish to disentangle the permeability decrease due
to volume reduction of the fracture from that due to low-velocity zones 
induced by surface fluctuations.  The first step is to
consider ``zero-mean'' self-affine surfaces, whose mean height $h$ 
is constant, but which have a tunable roughness.  The previous profile,
$z(x)$ for $0<x<L$, is rescaled so as to have unit variance, and then
multiplied by an adjustable amplitude $A$, giving adjustable variance $A^2$.
If we assume that the effect of the boundary perturbation is to give low
velocity zones which do not contribute to the flow, and that the volume of
these zones contributing to the dissipation integral is proportional to $A$, 
we would write in analogy to Eq.~\ref{rate} 
\begin{equation}
\label{rate2}
Q = Q_0 \left[ 1 - \frac{3 C_2 A}{h} \right] 
\end{equation}
Here $A$ is analogous to $R_1~L^{\zeta}/2$, and $C_2$ an order-1 fitting
coefficient.  In Fig.~\ref{fig-zero-mean}
we show the results obtained by numerical simulation
corresponding to a system with $h=32$ and $L=64$ (and, to be specific, 
$\nabla p = 6.25\times 10^{-6}$, $\mu=0.1$, $Y=4$,
which gives $Q_0=0.683$).  The filled circles are the numerical results, and
we see the expected linear decrease of the permeability with $A$.
In fact, a linear fit to the numerical data gives
$Q_0 = 0.72 \pm 0.03$ and $C_2=0.77 \pm 0.1$, which
are reasonable numerical values given our assumptions.

Further evidence for our arguments concerning the effects of low-velocity 
zones can be adduced by showing that the permeability is unchanged if they
are deleted.  We begin with the family of adjustable-amplitude self-affine surface just discussed, and filter out
the smallest wavelengths from the spectrum so as to generate
a smother surface.  The smooth surface is then shifted upwards so as to 
enclose the deleted surface fluctuations, in the RMS sense.  Quantitatively, 
if we have $N_L$ points on the $x$-axis, with $L=N_L \Delta$, and filter 
all wavelengths smaller than $\lambda_c = \Delta N_L/N_C$, then
\begin{equation}
\label{dh}
(\Delta h)^2 = \frac{1}{(N_L-N_C)^2}
\sum_{n=N_C+1}^{N_L} |Z(k_n)|^2  \qquad k_n = \frac{2\pi n}{\Delta N_L}
\end{equation}
The prefactor on the right arises from the discrete form of Parseval's
theorem \cite{recipes}, and in practice we choose $\Delta=1$.
We then simulate fluid flow in the smoothed fracture,
and in Fig.~\ref{fig-zero-mean} the open squares show the results obtained
as a function of roughness amplitude $A$.  The agreement between 
the actual and filtered surfaces is clear.

The extension of this analysis to the 3-D case is conceptually
straightforward, since the decrease in the permeability in this limit just
corresponds to the decrease in pore space {\em volume}.
The latter would be $L_x \times L_y \times R$, due to the
self-affine topography of the fracture surface, and the relevant expansion
parameter is again $R/h_0$.  The effects of low-velocity regions is
presumably the same as well.


\section{Narrow fractures}
\label{vertical}

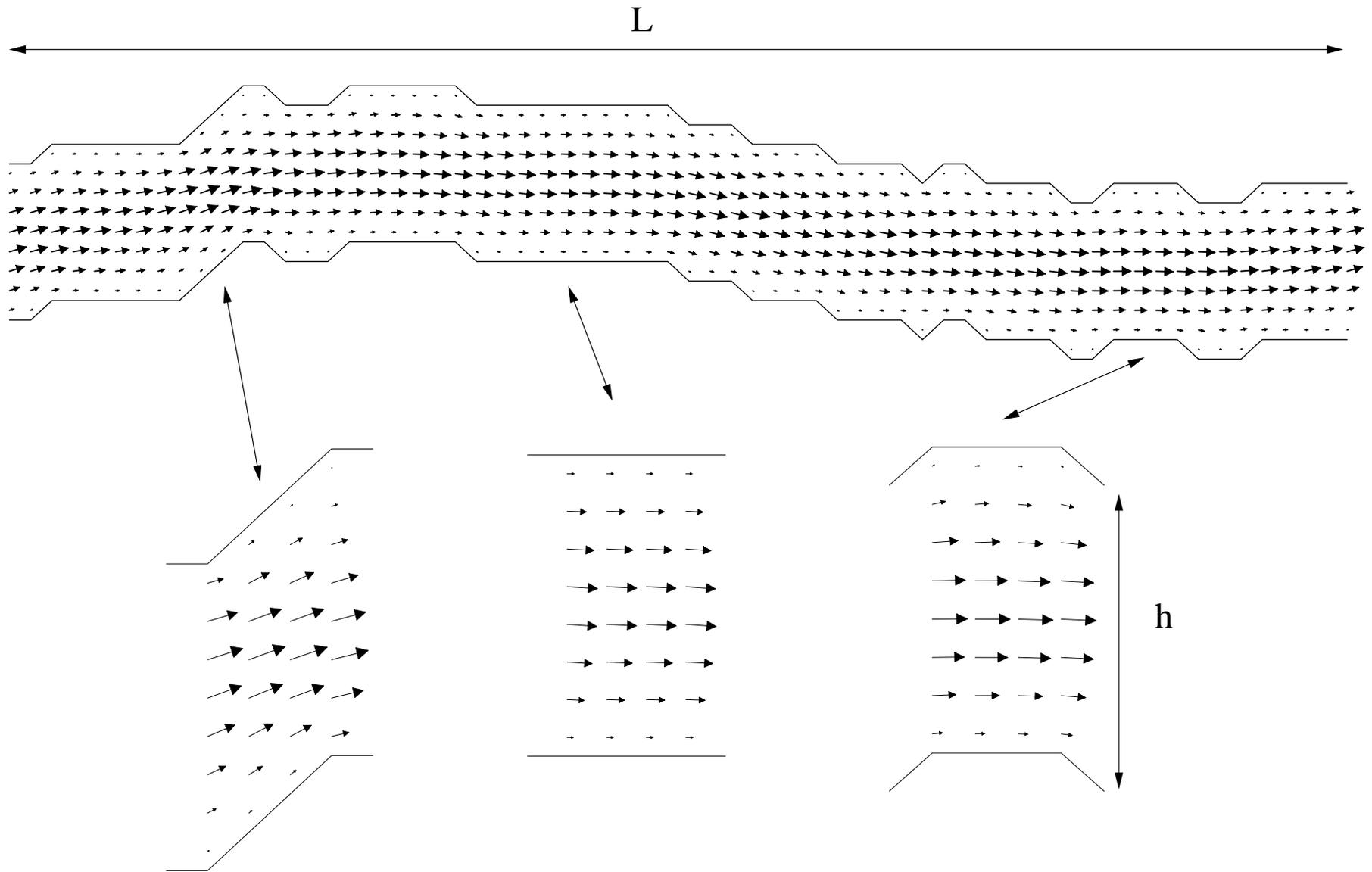
\begin{figure}
\caption{Flow field in a narrow trough fracture with a constant gap,
and exponent $\zeta=0.80$. The enlargements illustrate the
effect of the effective aperture on the flow field. }
\label{displacement}
\end{figure}

We now turn to the complimentary situation of narrow fractures, where the 
pore space is winding and convoluted.  
Consider the situation in which a rock is fractured and the 
two surfaces are simply displaced by a distance $h \ll L$, perpendicular 
to the mean fracture plane, with no lateral shift.  The vertical aperture 
is constant everywhere, but the effective local aperture for fluid flow, 
{\em i.e.}, the local width of the channel normal to the mean flow direction,
strongly depends on the local angle between the surface and the mean plane.
In Fig.~\ref{displacement} we show a fracture of length $L=64$,
generated by a self-affine surface of exponent $\zeta=0.8$, separated a 
distance $h=8$, along with the (lattice Boltzmann) computed velocity field.

The theoretical analysis will be based on a kind of local lubrication
approximation, where the channel is divided into a sequence of quasi-linear
segments at varying orientation angles.   First, we 
estimate the length $\xi_{\parallel}$ in the direction 
of the mean flow over which the channel formed by 
the two fracture surfaces can be considered approximately straight, which is
a typical size over which the fluctuations in the vertical direction
are small compared to the effective aperture of the channel.  Using the
self-affine scaling law for the correlation function,
\begin{equation}
\label{variance}
\sigma_z^2(\xi_\parallel) =
\langle [z(x,y)-z(x+\xi_{\parallel},y)]^2 \rangle = \phi(\ell) 
\left({ \xi_\parallel\over\ell} \right)^{2\zeta}
\end{equation}
where $\ell$ is a microscopic length, say a grain size, and $\phi(\ell)$
is then on the order of a grain size squared, we estimate
\begin{equation}
z(x,y)-z(x+\xi_{\parallel},y) 
\sim \ell \left({ \xi_\parallel\over\ell} \right)^\zeta 
\nonumber
\end{equation}
A segment of length $\xi_\parallel$ is roughly straight when this quantity 
is a small fraction of $h$, which yields
\begin{equation}
\label{delta-h}
\xi_\parallel \sim \ell \left({ h \over\ell }\right)^{1/\zeta}
\end{equation}
 
Returning to the entire fracture, 
each $\xi_{\parallel}$-channel is oriented at some angle $\theta_i$ 
with respect to the mean plane, and has effective aperture 
$a_i=h \cos\theta_i$, and length 
$\xi_{\parallel}^i=\xi_{\parallel}/\cos\theta_i$. 
Assuming Poiseuille flow in each $\xi_{\parallel}$-channel, the 
corresponding local permeability is $k_i=a_i^2/12$. 
We compute the pressure drop across the fracture by adding the 
the local pressure drops along the straight $\xi_{\parallel}$-channels 
along the whole path.  Noting that the flow rate $Q=\bar{U} h = \bar{U}_i 
a_i$ is constant,
\begin{eqnarray}
\Delta P &=& \sum_{i=1}^N \Delta P_i = 
- \sum_{i=1}^N \left( \frac{\mu}{k_i} \bar{U}_i \right) \xi_{\parallel}^i = 
\\ \nonumber
&=& - \sum_{i=1}^N 
\left(\frac{12 \mu}{h^2 \cos^2\theta_i} \right)
\left( \frac{Q}{h \cos\theta_i} \right)
\left( \frac{\xi_{\parallel}}{\cos\theta_i} \right)
\\ \nonumber 
&=& - \frac{12 \mu \xi_{\parallel} Q}{h^3} \sum_{i=1}^N \cos^{-
4}(\theta_i)
= - \frac{12 \mu \xi_{\parallel} \bar{U}}{h^2} \sum_{i=1}^N \cos^{-
4}(\theta_i) 
\nonumber
\end{eqnarray}
Finally, noting that $N=L/\xi_{\parallel} \gg 1$ is the number of channels,
we can convert the sum into an average over the distribution of angles, 
and write for the permeability
\begin{equation}
\label{k-h}
k = \frac{h^2}{12} \frac{1}{\langle \cos^{-4}(\theta) \rangle}
\end{equation}
 
We can give a simple if heuristic estimate of the permeability as follows.
Since the exponent $\zeta < 1$, the channels have small vertical fluctuations,
and we can approximate the cosine as
\begin{equation}
\cos\theta = \frac{\xi_{\parallel}}{\sqrt{\xi_{\parallel}^2 + 
\sigma_z^2(\xi_{\parallel})}} \approx 
1- \frac{1}{2} \left(\frac{\sigma_z(\xi_\parallel)}{\xi_\parallel}\right)^2
\end{equation}
Substituting in Eq.~\ref{k-h}, we obtain
\begin{equation}
\label{perm-narrow}
k \approx \frac{h^2}{12} 
\left[ 1 - 2 \left(\frac{\sigma_z(\xi_\parallel)}{\xi_\parallel}\right)^2
\right]
\end{equation}

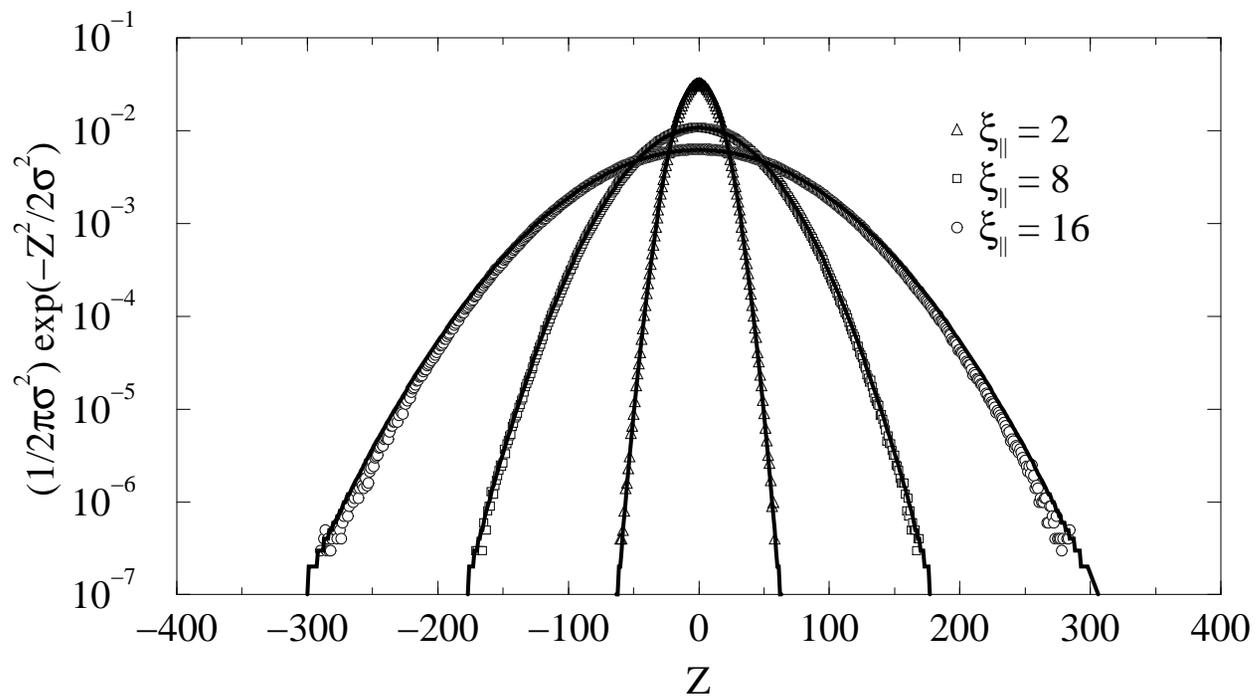
\begin{figure}
\caption{Top:  semi-log plot of the distribution of heights 
$Z=z(x)-z(x+\xi_{\parallel})$ for different values of $\xi_{\parallel}$ 
for a self-affine surface with exponent $\zeta=0.80$.  The solid lines 
are a Gaussian fit to the numerical results. 
Bottom:  variation of the mean spatial correlation 
$\langle Z^2 \rangle$ with $\xi_{\parallel}$, for $\zeta=0.80$ and 
$\zeta=0.95$. }
\label{G-distribution}
\end{figure}

A more convincing evaluation of the angular average makes use of the 
actual height distribution.   Experimental measurements indicate a Gaussian
distribution for the spatial correlation in heights,
$Z\equiv z(x,y)-z(x+\xi_{\parallel})$ \cite{plouraboue95}, and in fact our 
numerical procedure for generating self-affine surfaces also
gives a Gaussian distribution for $Z$.  We illustrate this point in
Fig.~\ref{G-distribution}, where we plot our generated probability
distribution function of $Z$, corresponding to different values of
$\xi_{\parallel}$, along with their Gaussian fits
\begin{equation}
p(Z) = \frac{1}{\sqrt{2\pi \sigma^2_z(\xi_{\parallel})}}
\exp{\left(-\frac{Z^2}{2 \sigma^2_z(\xi_{\parallel})}\right)}
\end{equation}
We also plot the (numerically-obtained) dependence of the mean spatial 
correlation, $\langle Z^2 \rangle = \sigma_z^2(\xi_{\parallel})$, on 
$\xi_\parallel$ for two choices of $\zeta$, confirming the scaling 
given above in Eq.~\ref{variance}.

The angular average is then given by
\begin{eqnarray}
\label{cosine}
\langle \cos^{-4}\theta \rangle &=& 
\int dZ\ p(Z) 
\left[
1 + 2 \left(\frac{Z}{\xi_{\parallel}}\right)^2 +
\left(\frac{Z}{\xi_{\parallel}}\right)^{4}
\right]
\\ \nonumber
&=&
1 + 2 \left(\frac{\sigma_z(\xi_\parallel)}{\xi_\parallel}\right)^2
+ 3 \left(\frac{\sigma_z(\xi_\parallel)}{\xi_\parallel}\right)^4
\end{eqnarray}
Using Eqs.~\ref{variance},\ref{k-h}, this result is consistent with 
the previous estimate Eq.~\ref{perm-narrow} based on simple scaling arguments.

Moreover, aside from the purely numerical coefficients, we can argue that 
scale invariance requires the angular average to have the form given in 
(\ref{cosine}).  The self-affine property of the fracture geometry implies 
that the probability distribution of heights $p(Z)$ should only
depend on the rescaled variable $Z/\xi_{\parallel}^\zeta$ 
\cite{plouraboue95}. More specifically, 
\begin{equation}
p(Z) = \sigma_z^{-1}(\xi_{\parallel})\; 
\psi[Z/\sigma_z(\xi_{\parallel})] 
\end{equation}
where the prefactor comes from the normalization.
If $\langle \cos^{-4}\theta \rangle$ is evaluated for this distribution,
we obtain a variant of Eq.~\ref{cosine} where the numerical coefficients 
2 and 3 are replaced by the moments of the function $\psi$, leading to
the same scaling form for the permeability.

Finally then, using the leading term in Eq.~\ref{cosine} for the angular 
average,
and Eq.~\ref{delta-h} for the value of $\xi_\parallel$, we have the result 
for the permeability of a narrow two dimensional self-affine fracture, 
\begin{eqnarray}
\label{perm-final}
k =  \frac{h^2}{12} \left[ 1- 2 
\left( \frac{\phi(\ell)}{\ell^2} \right)
\left( \frac{h}{\ell} \right)^{\frac{2\zeta-2}{\zeta}} \right] 
\end{eqnarray}
The principal approximation used in obtaining this result is that 
the fracture aperture may be regarded as a sequence of almost-straight 
segments.  
  
To test Eq.~\ref{perm-final} numerically, we first recast it in terms of the
fluid flux.  For a straight channel of height $h$, width $Y$ and length $L$,
and pressure drop $P$, we have flux $Q_0=h^3PY/12\mu L$.  Thus
\begin{equation}
Q_0 - Q \approx C_3 \frac{PY}{6\mu} ~ {\ell}^{\frac{2-2\zeta}{\zeta}} ~
h^{\frac{5\zeta-2}{\zeta}}
\label{flow-disp}
\end{equation}
where $C_3$ is another adjustable parameter expected to be of order one,
and we have taken $\ell$ to be the (unit) lattice spacing.

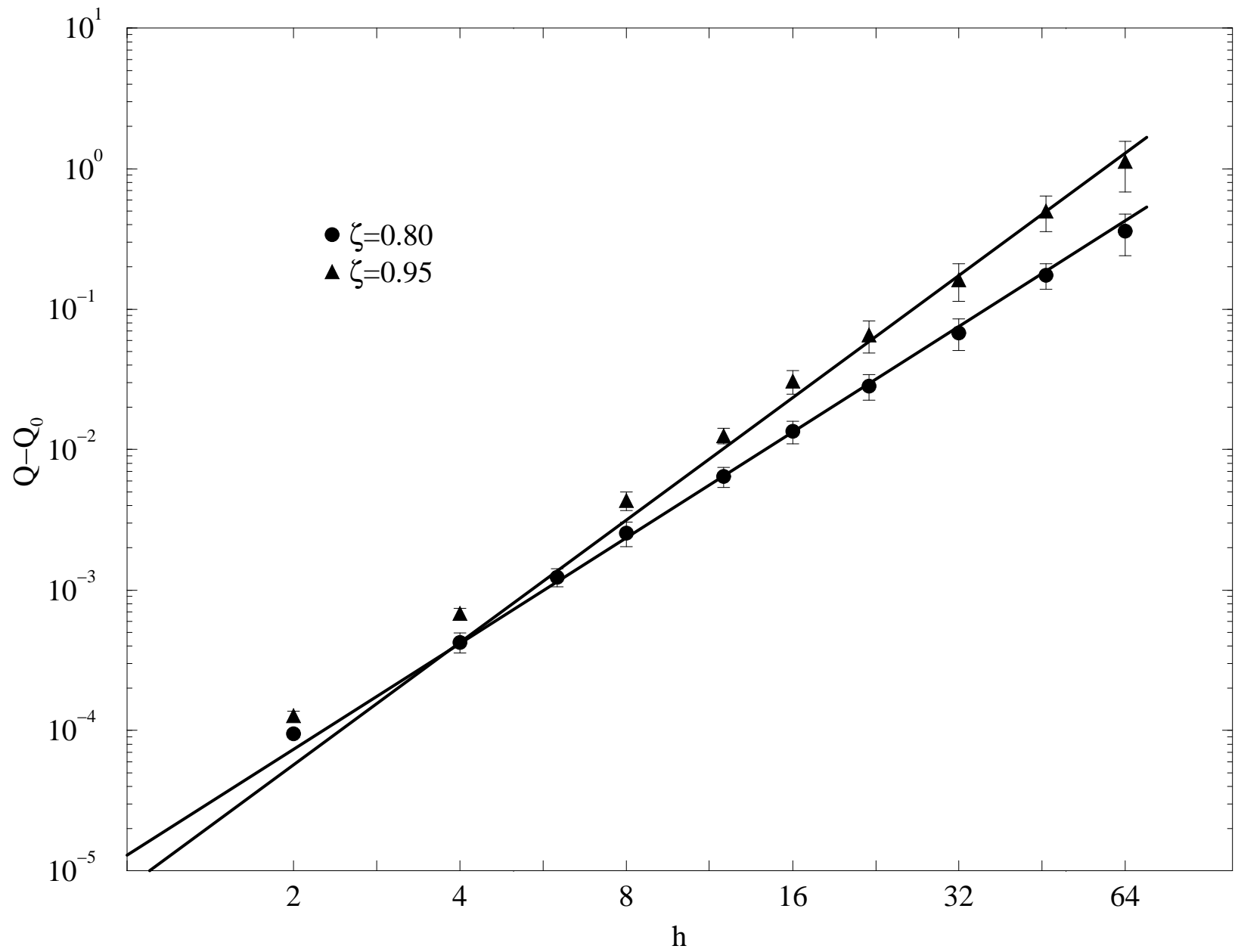
\begin{figure}
\caption{Change in flow rate as a function of gap $h$ 
between vertically-displaced self-affine surfaces, 
for exponents $\zeta=0.80$ and $\zeta=0.95$ and length $L=128$. }
\label{fdisplacement}
\end{figure}

In Fig.~\ref{fdisplacement}
we present the correction to $Q_0$ as a function of the 
distance between the opposite surfaces $h$, for both roughness
exponents $\zeta=0.80$ and $\zeta = 0.95$.
We find the good agreement for the predicted exponent
for the case $\zeta=0.80$. The only 
adjustable parameter is the coefficient, which is found to be $C_3 = 3.1 
\pm 1.2$ in fairly good agreement with the expected value. The fitted 
exponent is $2.43 \pm 0.08$, which agrees with the predicted value 
$(5\zeta-2)/\zeta = 2.5$. 
On the other hand, in the case $\zeta=0.95$ the numerical results 
deviates from the asymptotic behavior at small distances 
between the surfaces. That is because the successive powers given
by the expansion of Eq.~\ref{perm-final} are very close to
each other (the difference between them is 
$(2\zeta-2)/\zeta \approx 0.1$).  Again if only the coefficient is
fitted, we find 
$C_3 = 0.35 \pm 0.21$. Fitting also the exponent,
but using only the numerical simulations with large separation between
surfaces, we obtain $2.71 \pm 0.39$ where $(5\zeta-2)/\zeta = 2.89$.

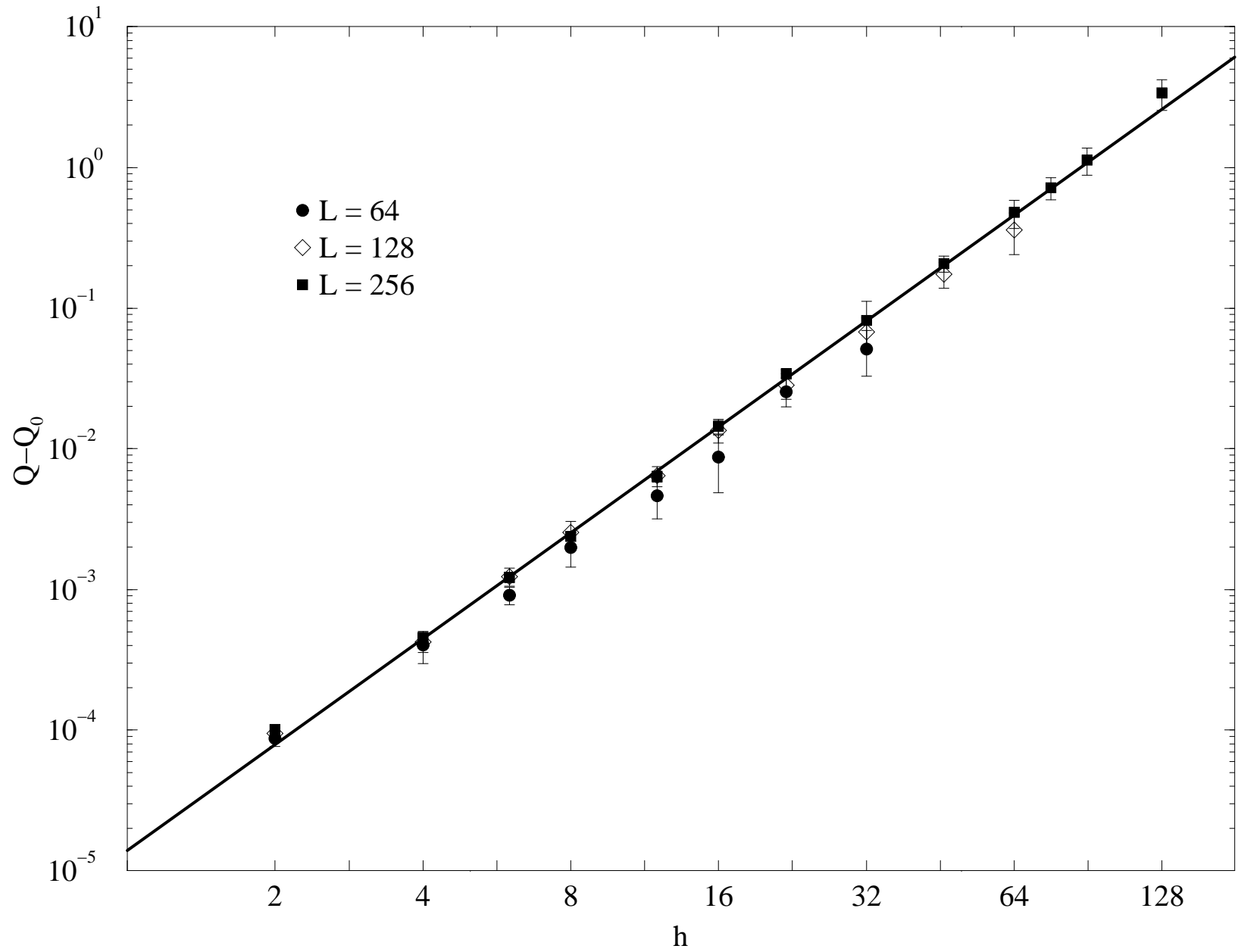
\begin{figure}
\caption{Change in the flow rate as a function of gap $h$ 
for $\zeta=0.80$ and various $L$.  The solid line is a fit to the largest
($L=256$) system.}
\label{size-disp}
\end{figure}

In Fig.~\ref{size-disp} we compare numerical results corresponding
to systems of different sizes with the same roughness exponent
$\zeta=0.80$. It can be seen that the accuracy in the exponent is
improved as the size of the system is larger. 
We obtain a value $2.52 \pm 0.04$ for the largest system
with size $L=256$.

As mentioned, the crossover between small and large surface 
roughness, (compared to the mean aperture of the fracture), can be 
addressed numerically. The narrow fracture regime, where
$Q_0-Q \propto h^{\frac{5\zeta-2}{\zeta}}$, is expected to be 
valid when $\xi_{\parallel} \ll L$. The critical mean aperture value $h_c$
which corresponds to $\xi_{\parallel} \sim L$ turns out to be, 
using the fitted value of $C_3$, $h_c \sim 100$. On the other hand, the 
small surface-deviation scaling behavior can be obtained from
Eq. (\ref{rate}), where $L$ is the constant size of the system and
$h_0$ is now the variable mean aperture of the fracture $h$,
\begin{equation}
\label{small}
Q_0-Q \approx \frac{P Y}{12 \mu} \frac{3 C_1 R_1 L^{\zeta}}{2} h^2
\end{equation}

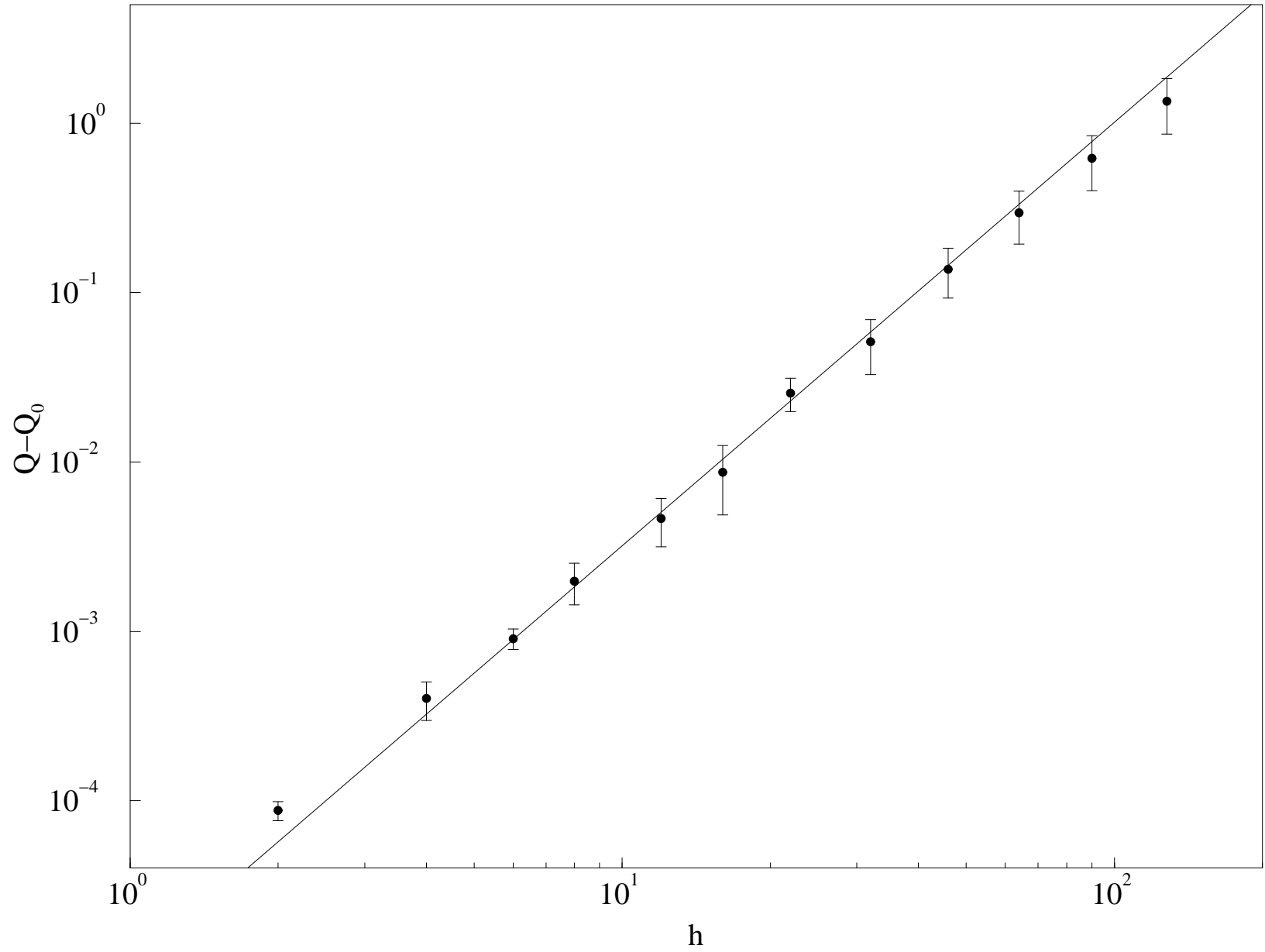
\begin{figure}
\caption{Change in the flow correction as a function of the
gap $h$ for $\zeta=0.80$ and $L=64$. The solid line is the 
to the narrow fractures regime. At large gaps ($h \ge 64$)
a small deviation towards a smaller exponent can be observed}
\label{crossover}
\end{figure}

In Fig.~\ref{crossover} we display the correction to $Q_0$ in the range
$2 \leq h \leq 128$ for a system size $L=64$. A small deviation from
the narrow fractures regime can be observed, starting at $h=64$. 
As the size of the gap becomes larger, the correction to the unperturbed
flow rate $Q_0$ decreases as a power of $h$. Therefore, both
larger computational time (starting the simulation from an initially at rest fluid
the time interval to reach the steady asymptotic value grows as the square
of the gap size \cite{batchelor}) and higher accuracy in the 
fluid field close to the solid surface are needed. Our current
computational limitations prevents us to
simulate systems where $L \gg \xi_{\parallel}$. Nevertheless, 
we believe that the observed deviation correspond to a crossover towards 
a smaller exponent ($h^{2}$) 
as expected from the previous discussion.

%
%

\section{Fractures with shifted surfaces}
\label{parallel}

In the field, when a rock is fractured its two sides may be shifted
laterally parallel to the mean plane by geological processes.  We now
consider how our results for permeability are modified by this effect.
The situation is somewhat different in two dimensions
than in three.  In order that a two dimensional fracture remain open it
is necessary that the two sides do not touch, whereas in three dimensions 
in the presence of a lateral shift it is easy for the two surfaces to touch at
one or a few points in the plane, while the aperture is still open to flow.
Furthermore, in two dimensions when the fracture nearly touches at a point,
the permeability will be dominated by the large pressure drop needed to force
fluid through this narrow gap.  In fact, in this case an estimate of the
permeability is simply $k\approx a_{\rm min}^2/12$, where $a_{\rm min}$ is
the minimum value of the aperture.  In this section we address the
complimentary case where the fracture is distinctly open.
 
As discussed in Section 2, in the presence of a lateral shift $d$, 
the fracture aperture is now 
the random function $a_d(x,y)$ given by Eq.~\ref{aperture} rather than the
constant $h$.  However, there is a residual correlation between the two sides 
of the fracture, which allows us to relate the permeability to the
self-affine statistics.  First, we determine a condition on the shift for the
fracture to be open.  Provided the average properties of the surface are
statistically stationary (independent of $x$), we have 
$\langle a_d(x)\rangle = h$, and the variance of the aperture distribution is
\begin{equation}
\sigma_a^2 (d) \equiv \langle [ a_d(x) - \langle a_d(x) \rangle]^2 \rangle =
\langle [z(x+d)-z(x)]^2 \rangle = \sigma_z^2(d) = \phi(\ell) 
\left( {d\over \ell} \right)^{2\zeta}
\end{equation}
The aperture is surely open when $\sigma_a(d) \ll h$, which gives
$d \ll \ell (h/\ell)^{1/\zeta}$, but from Eq.~\ref{delta-h}, this is 
just $d \ll \xi_\parallel$.  In other words, the shift must be small compared
to the typical length of a straight segment of unshifted channel, and there
is little change in the geometry compared to the unshifted case.
We expect, therefore, that the previous result Eq.~\ref{flow-disp} should apply
at least up to a different value of the numerical coefficient $C_3$.

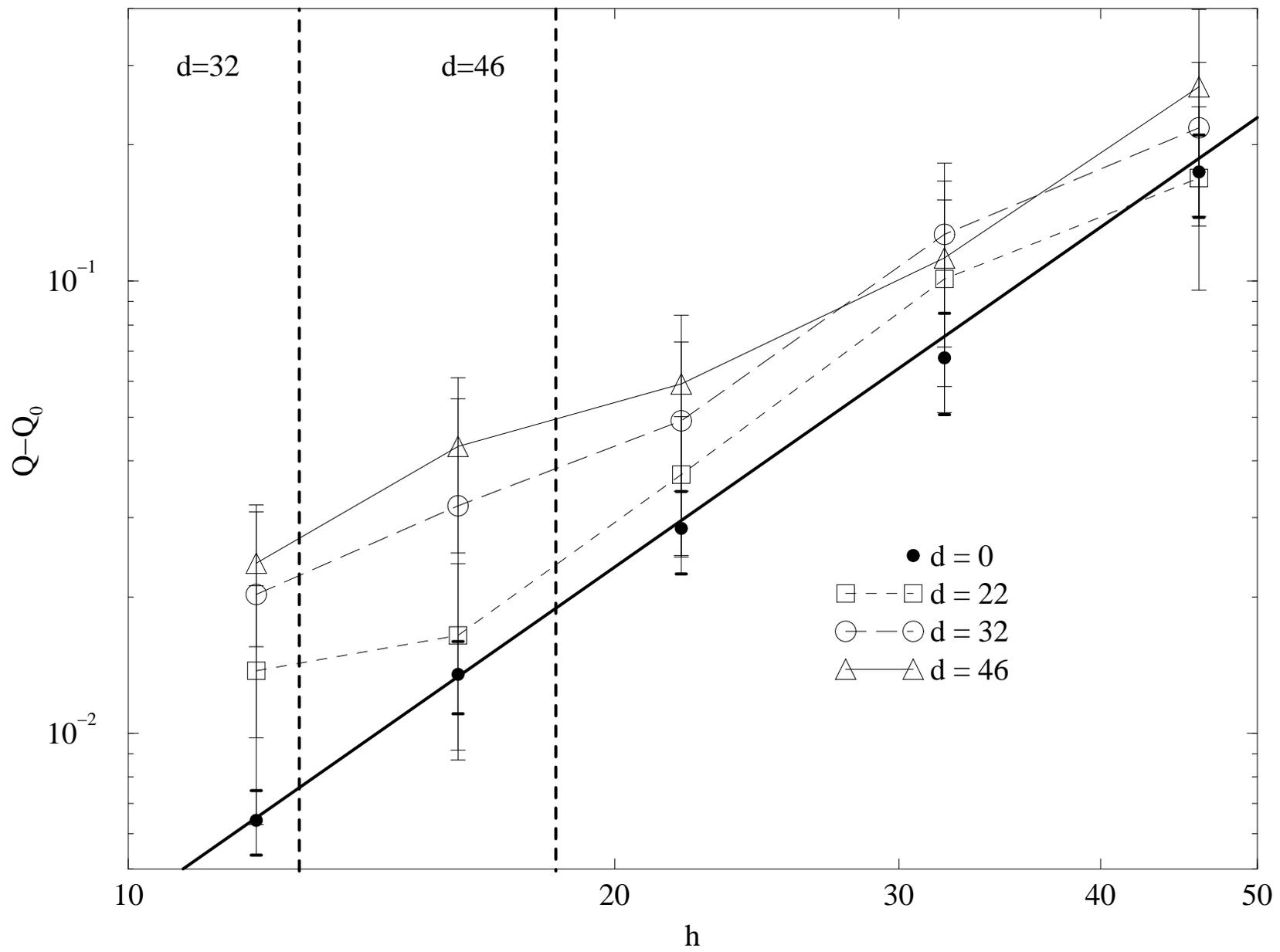
\begin{figure}
\caption{Change in flow rate correction as a function of gap $h$
for exponent $\zeta=0.80$ and various lateral shifts $d$.
The solid line is a fit to the case $d=0$ shown in 
Figure~\ref{fdisplacement}. The two vertical dashed lines
divide the regions $(h/\ell)<(d/\ell)^{\zeta}$ and 
$(h/\ell)>(d/\ell)^{\zeta}$, for $d=22$ and $d=46$.}
\label{fshift}
\end{figure}

In Fig.~\ref{fshift} we show LB simulation results for the correction to the 
unperturbed flow rate $Q_0$, for different values of the shift $d$.
We see that the behavior is consistent with these arguments
for all values of the shift, with deviations occurring when the shift is 
too large for the mean aperture.  Several other shift values were
simulated obtaining the same general behaviors as in those shown in
Fig.~\ref{fshift}. As is often the case with 
this type of scaling argument, reasoning in terms of asymptotic limits
leads to conditions that one quantity must be {\em much} larger than 
another, but in practice the range of validity is much wider.

\section{Conclusions}

We have studied the permeability of two-dimensional self-affine fractures,
using asymptotic analytic arguments for wide and narrow apertures.
Numerical simulations using the lattice Boltzmann method have verified the  
predictions, and also suggest a smooth crossover 
between the limits used in the derivations.
We have obtained expressions for the permeability in which the usual 
expression for straight channels is modified by terms related to the
Hurst exponent characterizing the fracture surface.  

In the wide-gap case, we obtained a perturbative correction related to the
roughness exponent and amplitude.  The result is formally identical to 
the lubrication theory prediction, but the corrections are known, in
principle, and furthermore we can understand the discrepancies between 
the derived numerical coefficient and simulation in
terms of the effects of low-velocity zones which do not contribute to
transport.  In the narrow gap case, 
we also derive an analytic correction related to the roughness
exponent and amplitude (or microscopic length $\ell$).  
The analysis is based in a local lubrication approximation,
dividing the whole fracture into a chain of approximately
linear channels with varying orientation angles respect to the 
main flow direction. The results is equivalent to a correction due to
the tortuosity of the fracture, where the tortuosity itself depends on the
gap size. Finally, we show that 
in the case when there is a lateral shift between the surfaces,
there is no qualitative change compared to the
unshifted case. This is basically due to the two dimensional geometry,
where in order to have an open channel the shift must be small
compared to the size of the typical length of a quasi-linear segment
of unshifted fracture.

The two-dimensional nature of the problem
imposes restrictions on the range of validity of analytic estimates, but 
the numerical results are in general agreement over a much wider range. 

The extension of this work to fully three-dimensional fractures is in 
progress.  The case of a wide gap does not require any further significant 
conceptual effort.  However, the narrow gap case the fracture geometry is
significantly more complicated because it is not feasible to use any
quasi-one-dimensional approximation for the flow paths.  Numerical
simulations are certainly feasible, at least up to size limits imposed by
one's computational hardware, but additional ideas are required for 
analytic arguments. Future work will consider diffusion, and also
convection plus diffusion, in these self-affine fractures, in both
two and three dimensions. The LB method is quite simple for numerical
simulations, and we will explore the way in which the roughness exponent
enters into the quantitative behavior,  
along the lines given here.

\section*{Acknowledgments}

We thank J. P. Hulin, F. Plourabou\'e and M. Tanksley for discussions.
This research was supported by the Geosciences Research Program, Office of
Basic Energy Sciences, U.S. Department of Energy, and computational facilities 
were provided by the National Energy Resources Scientific 
Computer Center.


\newpage
\begin{figure}
\caption{Example of the geometry and velocity field in a fracture with one 
self-affine surface of roughness exponent $\zeta=0.80$.
The enlargements show the difference in the velocity decay near smooth and
rough boundaries.}
\label{fracture}
\end{figure}

\begin{figure}
\caption{Perturbation in the flow rate in a wide, rough channel as a 
function of the length $L$ of the system for $\zeta=0.80$ and 
$\zeta=0.95$.} 
\label{q-vs-size}
\end{figure}

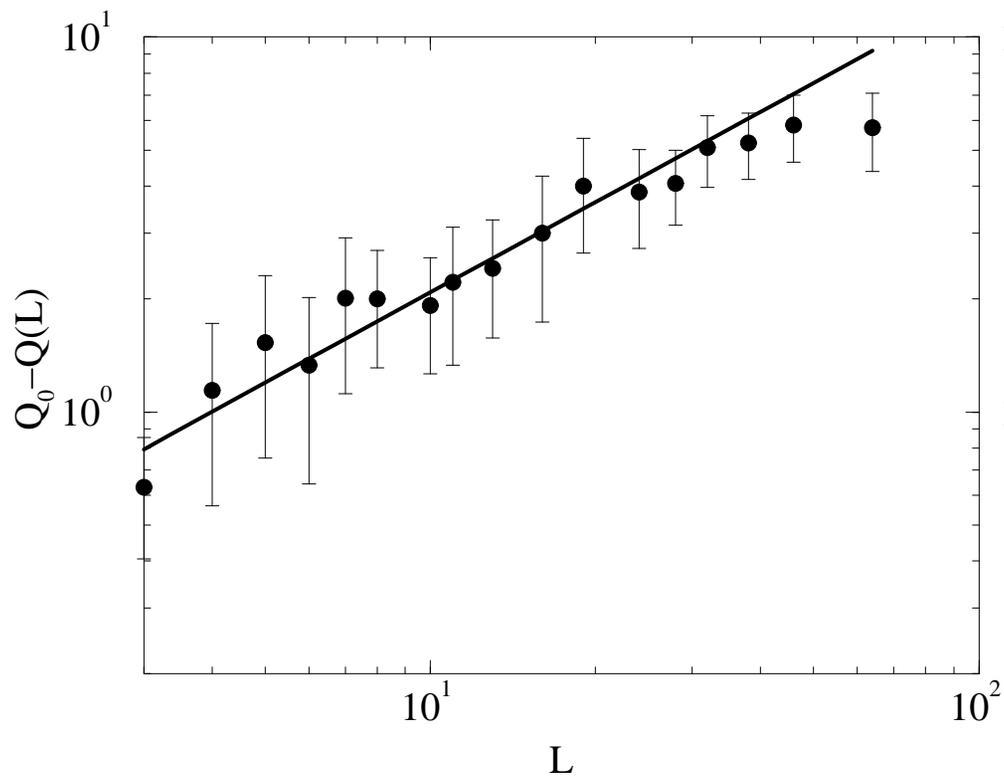
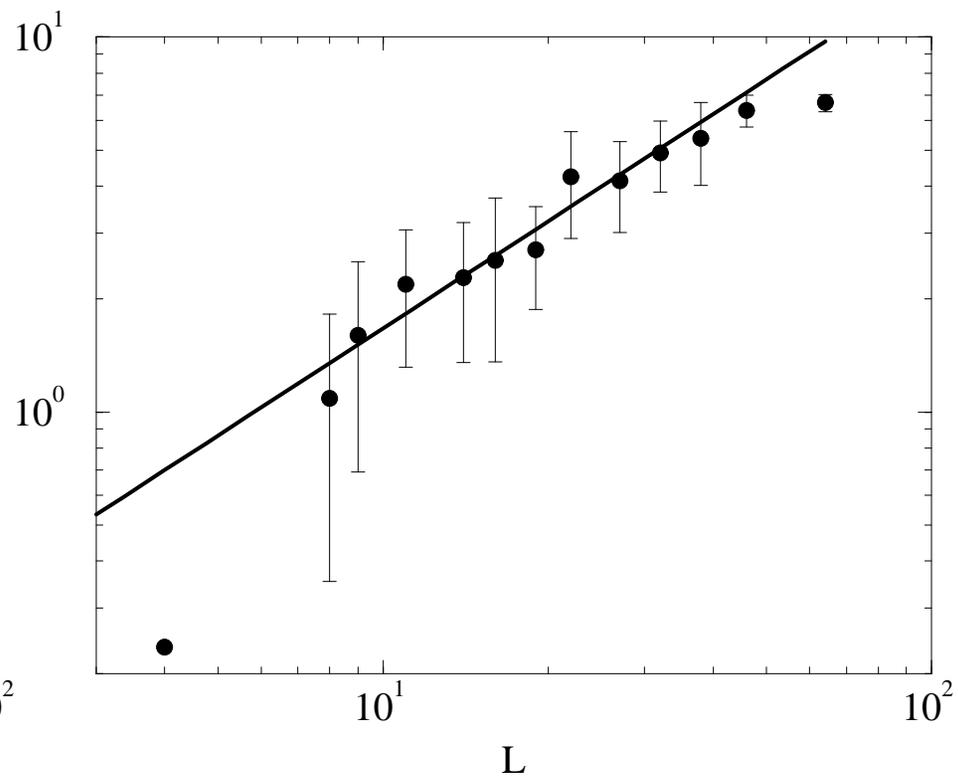
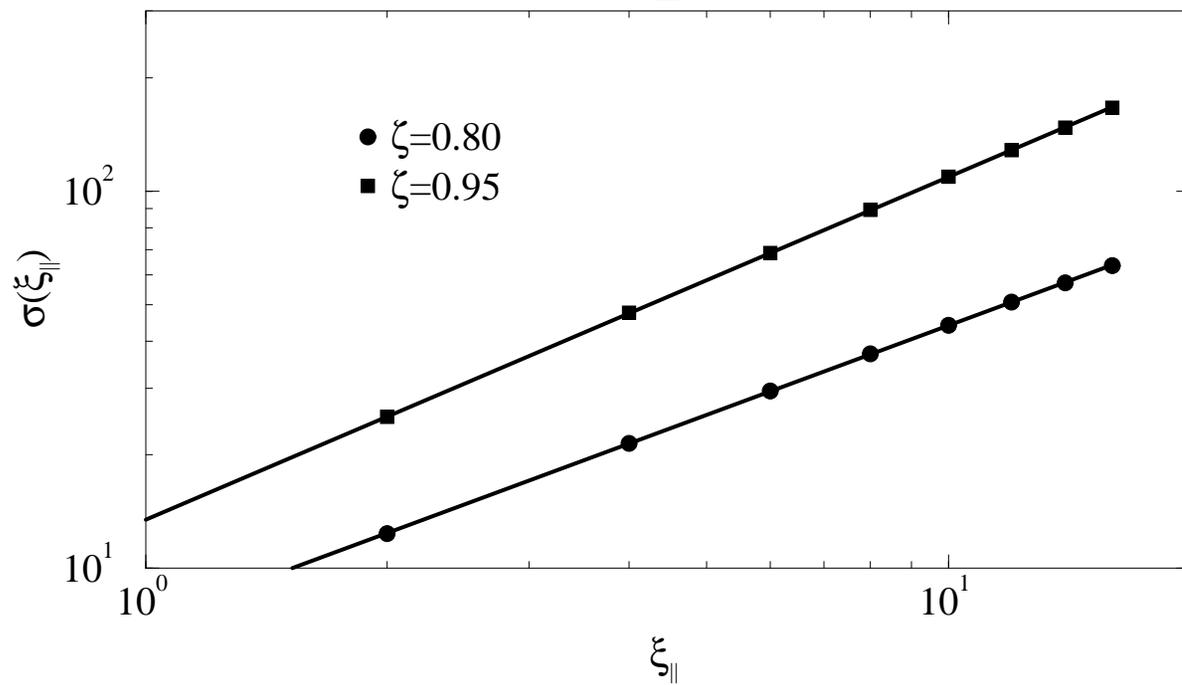
\end{document}